%% file: main.tex
\documentclass[twocolumn,10pt]{article}
\usepackage[top=.5in, bottom=.75in, left=.5in, right=.5in]{geometry}
\usepackage[utf8]{inputenc}
\usepackage{graphicx}
\usepackage{stfloats}
\graphicspath{{figures/}}
\usepackage{amsmath}
\newcommand\MyLBrace[2]{%
\left.\rule{-10pt}{#1}\right\}\text{#2}}
\usepackage{breqn}
\usepackage{amsfonts}
\usepackage[version=4]{mhchem}
\usepackage{siunitx}
\usepackage{longtable,tabularx}
\usepackage{multirow}
\usepackage{tikz}
\usepackage[export]{adjustbox}
\usepackage{blindtext}
\usepackage{float}
\usepackage{tablefootnote}
\usepackage{color}
\usepackage{mathtools}
\usepackage{multirow}
\usepackage{hyperref}
\usepackage[switch]{lineno}

\hypersetup{
pdftitle={A Probabilistic Graphical Model Foundation for Enabling Predictive Digital Twins at Scale},
pdfsubject={Digital Twin},
pdfauthor={Kapteyn, M.G., Pretorius, J.V.R., and Willcox, K.E.},
pdfkeywords={Digital Twin, Uncertainty Quantification, Probabilistic Graphical Model, Self-Aware Vehicle}
}

\gdef\notesoff{\gdef\note##1{}}
\notesoff

\usepackage{titlesec}
\titlespacing*{\section}{0pt}{0.6\baselineskip}{0.3\baselineskip}

\DeclareMathOperator*{\argmax}{arg\,max}

\title{\huge{A Probabilistic Graphical Model Foundation for\\Enabling Predictive Digital Twins at Scale}}
\author{\normalsize{Michael G. Kapteyn$^{1}$\:, Jacob V.R.\ Pretorius$^{2}$\:, and Karen E. Willcox$^3$\footnote{Corresponding author (kwillcox@oden.utexas.edu).}}\\
\footnotesize{$^1$ Department of Aeronautics and Astronautics, Massachusetts Institute of Technology, Cambridge, MA 02139, United States}\\
\footnotesize{$^2$ The Jessara Group, Austin, TX 78704, United States}\\
\footnotesize{$^3$ Oden Institute for Computational Engineering and Sciences, University of Texas at Austin, Austin, TX 78712, United States}
}
\date{}

\begin{document}
\maketitle

\vspace{-.3in}
\input{sections/0_abstract}
\input{sections/1_introduction}

\input{sections/2_results}
\input{sections/3_discussion}
\input{sections/4_methods}
\input{sections/9_acknowledgments}
\bibliographystyle{unsrturl}
\bibliography{references}

\end{document}

%% file: sections/0_abstract.tex
\begin{abstract}
\noindent
A unifying mathematical formulation is needed to move from one-off digital twins built through custom implementations to robust digital twin implementations at scale. This work proposes a probabilistic graphical model as a formal mathematical representation of a digital twin and its associated physical asset. We create an abstraction of the asset-twin system as a set of coupled dynamical systems, evolving over time through their respective state-spaces and interacting via observed data and control inputs. The formal definition of this coupled system as a probabilistic graphical model enables us to draw upon well-established theory and methods from Bayesian statistics, dynamical systems, and control theory. The declarative and general nature of the proposed digital twin model make it rigorous yet flexible, enabling its application at scale in a diverse range of application areas. We demonstrate how the model is instantiated to enable a structural digital twin of an unmanned aerial vehicle (UAV). The digital twin is calibrated using experimental data from a physical UAV asset. Its use in dynamic decision making is then illustrated in a synthetic example where the UAV undergoes an in-flight damage event and the digital twin is dynamically updated using sensor data.
The graphical model foundation ensures that the digital twin calibration and updating process is principled, unified, and able to scale to an entire fleet of digital twins.
\end{abstract}

%% file: sections/1_introduction.tex
\section*{Introduction}\label{sec:introduction}
A digital twin is a computational model (or a set of coupled computational models) that evolves over time to persistently represent the structure, behavior, and context of a unique physical asset such as a component, system, or process~\cite{aiaaDEIC}. The digital twin paradigm has seen significant interest across a range of application areas as a way to construct, manage, and leverage state-of-the-art computational models and data-driven learning. Digital twins underpin intelligent automation by supporting data-driven decision making and enabling asset-specific analysis. Despite this surge in interest, state-of-the-art digital twins are largely the result of custom implementations that require considerable deployment resources and a high level of expertise. To move from the one-off digital twin to accessible robust digital twin implementations at scale requires a unifying mathematical foundation. This paper proposes such a foundation by drawing on the theoretical foundations and computational techniques of dynamical systems theory and probabilistic graphical models. The result is a mathematical model for what comprises a digital twin and for how a digital twin evolves and interacts with its associated physical asset. Specifically, we propose a probabilistic graphical model of a digital twin and its associated physical asset, providing a principled mathematical foundation for creating, leveraging, and studying digital twins.

Digital twins have garnered attention in a wide range of applications~\cite{rasheed2020digital}. Structural digital twins have shown promise in virtual health monitoring, certification, and predictive maintenance~\cite{tuegel2011reengineering,glaessgen2012digital,li2017dynamic,podskarbi2020}. In healthcare, digital twins of human-beings promise to advance medical assessment, diagnosis, personalized treatment, and \textit{in-silico} drug testing~\cite{bruynseels2018digital,rivera2019towards,barricelli2020human}. Similarly, digital twins of individual students offer a path to personalized education~\cite{yu2017towards}. At a larger scale, smart cities enabled by digital twins and Internet of Things (IoT) devices promise to revolutionize urban planning, resource allocation, sustainability and traffic optimization~\cite{mohammadi2017smart}. Although each of these applications has its own unique requirements, challenges, and desired outcomes, the mathematical foundation we develop in this work focuses on a common thread that runs throughout: dynamically updated asset-specific computational models integrated within the data-driven analysis and decision-making feedback loop.

\begin{figure*}[h]
\centering
\includegraphics[width=0.9\textwidth]{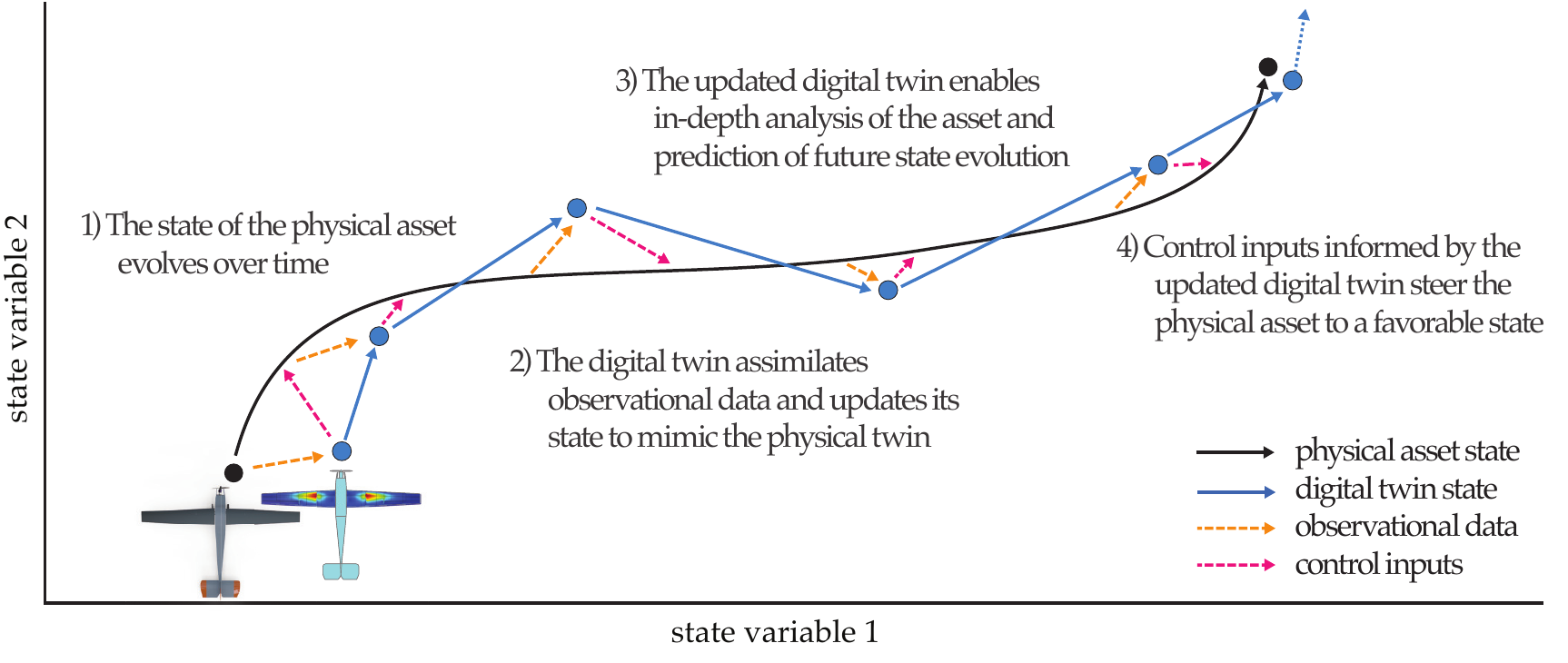}
\caption{Conceptual model of a physical asset and its digital twin, evolving over time through their respective state spaces. These two dynamical systems are tightly coupled: The digital twin uses observational data to estimate the state of the physical twin, so that it can, in turn, provide optimal control inputs that steer the physical asset to favorable states, while balancing other factors like maintaining observability over the asset state and minimizing control costs.}
\label{fig:conceptualmodel}
\end{figure*}

We adopt a view of the physical asset and its digital twin as two coupled dynamical systems, evolving over time through their respective state spaces as shown in Figure~\ref{fig:conceptualmodel}. The digital twin acquires and assimilates observational data from the asset (e.g., data from sensors or manual inspections) and uses this information to continually update its internal models so that they reflect the evolving physical system. This synergistic multi-way coupling between the physical system, the data collection, the computational models, and the decision-making process draws on the dynamic data driven applications systems (DDDAS) paradigm \cite{darema2004dynamic,blasch2018handbook}. The digital twin can then use these up-to-date internal models for analysis, prediction, optimization, and control of the physical system.

Motivated by this conceptual model, we develop a probabilistic graphical model~\cite{koller2009probabilistic} that defines the elements comprising this coupled dynamical system, and mathematically describes the interactions that need to be modeled in the digital twin. Our model draws inspiration from classical agent-based models such as the partially observable Markov decision process~\cite{russell2002artificial}, but includes important features that are unique to the digital twin context. The graphical model formalism provides a firm foundation from which to draw ideas and techniques from uncertainty quantification, control theory, decision theory, artificial intelligence, and data-driven modeling in order to carry out complex tasks such as data assimilation, state estimation, prediction, planning, and learning, all of which are required to realize the full potential of a digital twin.

Throughout this paper we demonstrate our proposed graphical model using a motivating application: the development of a structural digital twin of an unmanned aerial vehicle (UAV). In particular, we demonstrate the application of the probabilistic graphical model to two phases in the asset lifecycle. First, we show how experimental calibration can be used to transform a baseline structural model into a unique digital twin tailored to the characteristics of a particular asset. Here the proposed graphical model serves as a rigorous framework for deciding which calibration experiments to perform, leveraging the resulting experimental data for principled model calibration, and evaluating the performance of the calibrated digital twin. Next, we demonstrate how the calibrated structural digital twin enters operation alongside the physical asset where it assimilates sensed structural data to update its internal models of the vehicle structure. The dynamically updated models are used for analysis and evaluation of the vehicle's structural health and for decision-making. Formulation of this process via the proposed graphical model enables end-to-end uncertainty quantification and principled analysis, prediction, and decision-making.
%
%

This motivating example has specific application in a number of settings. One is urban air mobility and autonomous package delivery, where UAVs operate in urban environments and are subject to damage. In order to ensure a safe, robust, reliable and scalable system, these vehicles must be equipped with advanced sensing, inference, and decision capabilities that continually monitor and react to the vehicle's changing structural state. The vehicle uses this capability to decide whether to perform more aggressive maneuvers or fall back to more conservative maneuvers in order to minimize further damage or degradation. Another application is hypersonic vehicles, which operate in extreme environments and thus undergo continual degradation of their structural condition. More generally, the structural digital twin illustrative example is highly relevant to other applications where the condition of the system changes over time due to environmental influences and/or operating wear and tear. Examples include wind turbines, nuclear reactors, gas turbine engines, and civil infrastructure such as buildings and bridges.

%% file: sections/2_results.tex
\section*{Results}\label{chap:foundation}
We present three key results: the abstraction of a digital twin into a state-space formulation that is a basis for mathematical modeling, the realization of a digital twin mathematical model in the form of a dynamic decision network, and an application of the proposed model for the evolution of a UAV structural digital twin.

\subsection*{A Mathematical Abstraction of the Asset-Twin System}\label{sec:abstraction}
\begin{figure*}
\centering
\includegraphics[trim=0cm 0.5cm 0cm 0cm, width=0.89\linewidth]{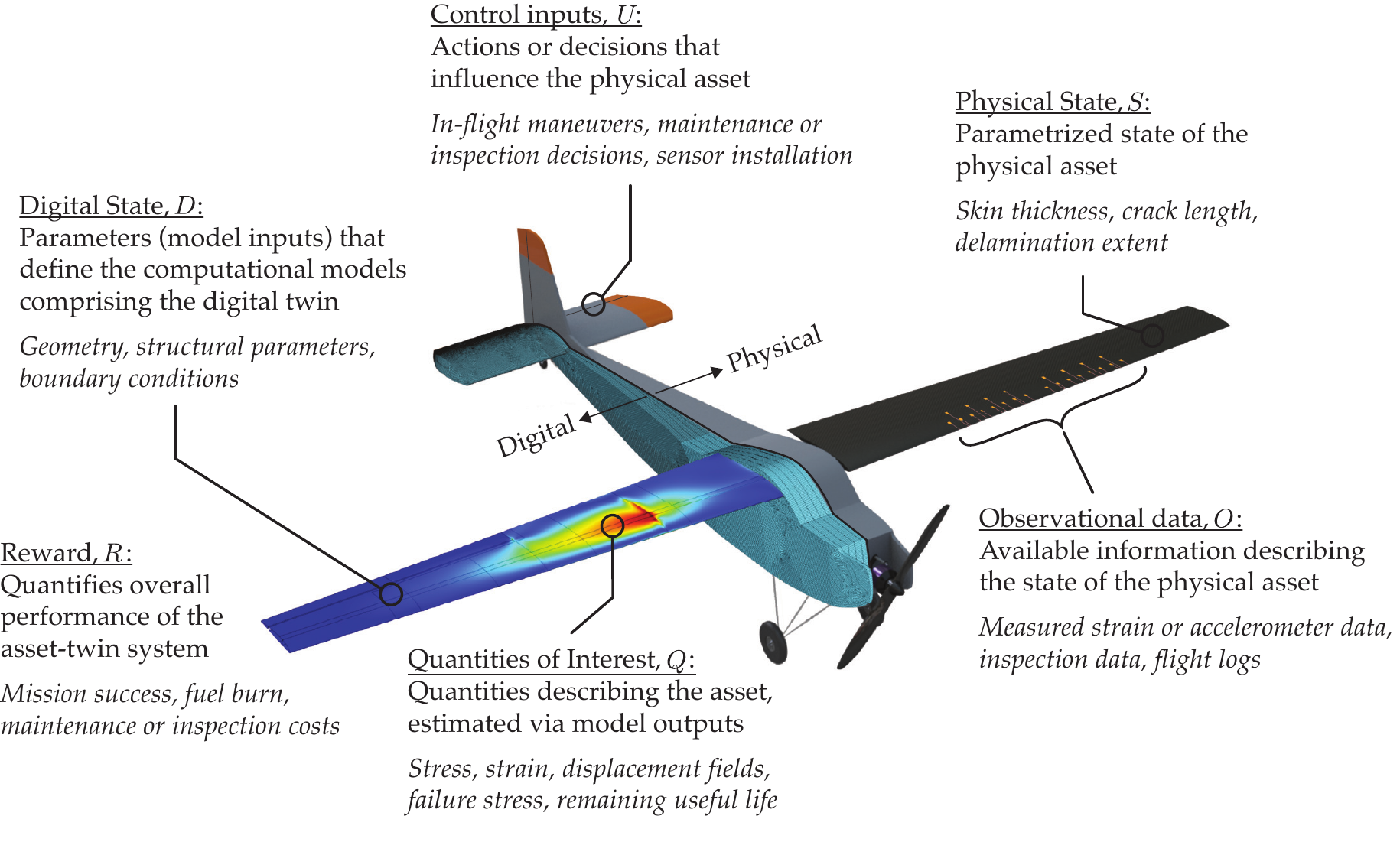}
\caption{Our abstraction of a digital twin and its associated physical asset. Specific examples from the self-aware aerial vehicle application are provided but the quantities themselves are abstract and can be used to model asset-twin systems in any application area.}
\label{fig:uavElements}
\end{figure*}

The first main result of this work is the abstraction of a digital twin and its associated physical asset into a representation comprised of six key quantities, defined in Figure~\ref{fig:uavElements}. We provide specific examples from the self-aware UAV application but emphasize that this abstraction can be applied to digital twins from any discipline and application, thus providing a unifying framework for describing and defining digital twins. Hence, the abstraction forms the basis for defining a general mathematical model for an asset-twin system, such as the probabilistic graphical model proposed in the following subsection.

Note the key differences between the physical and digital states. The physical state space encapsulates variation in the state of the asset and could thus be a complex high-dimensional space. The physical state is typically not fully observable. Together, these attributes make the physical state generally intractable to model directly. The digital state is defined as a set of parameters that characterize the models comprising the digital twin. The digital state is updated as the asset evolves over time or as new information about the asset becomes available. In defining the digital state, one must consider what information is sufficient to support the use case at hand. A well-designed digital twin should be comprised of models that provide a sufficiently complex digital state space, capturing variation in the physical asset that is relevant for diagnosis, prediction, and decision-making in the application of interest. On the other hand, the digital state space should be simple enough to enable tractable estimation of the digital state, even when only partially observable. For this reason, the digital state space will generally be only a subset of the physical state space.

\subsection*{A Probabilistic Graphical Model for Digital Twins}\label{sec:graphicalmodel}
The proposed abstraction of the digital twin leads to the second main result of this work, a probabilistic graphical model of a digital twin and its associated physical asset. This graphical model represents the structure in an asset-twin system by encoding the interaction and evolution of quantities defined in Figure~\ref{fig:uavElements}. In particular, the model encodes the end-to-end digital twin data-to-decisions flow, from sensing through inference and assimilation to action.

\begin{figure*}
\centering
\includegraphics[trim=0cm 0.5cm 0cm 0cm, width=0.89\linewidth]{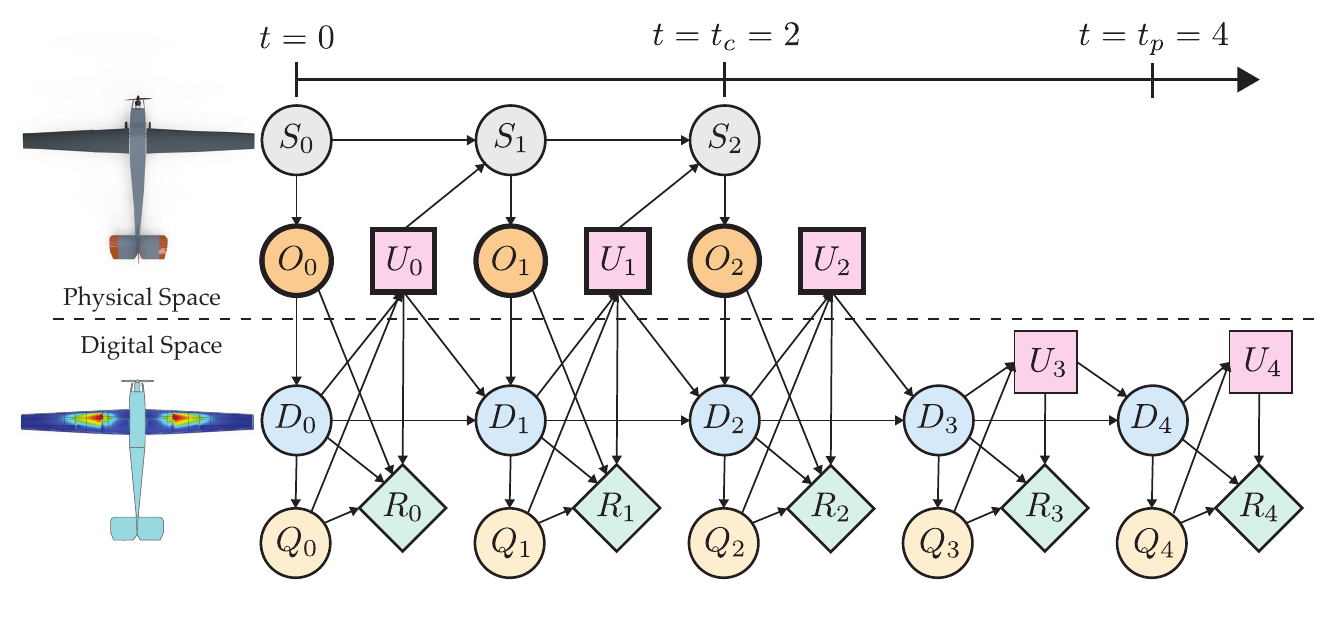}
\caption{Our approach results in a dynamic decision network that mathematically represents a physical asset and its digital twin. Nodes in the graph shown with bold outlines are observed quantities (i.e., they are assigned a deterministic value), while other quantities are estimated (typically represented by a probability distribution). Directed edges represent conditional dependence. Here the control nodes are decision nodes.}
\label{fig:graphicalmodel}
\end{figure*}

Formally, the model presented here is a dynamic decision network: a dynamic Bayesian network with the addition of decision nodes. Figure~\ref{fig:graphicalmodel} shows the graph unrolled from the initial timestep, $t=0$, to the current timestep, $t=t_c$, and into the future to the prediction horizon, $t=t_p$. Nodes in the graph are random variables representing each quantity at discrete points in time. Throughout this section we use upper-case letters to denote random variables, with the corresponding lower-case letter denoting a value of the random variable. For example, the digital state at timestep $t$ is estimated probabilistically by defining the random variables $D_t \sim p(d_t)$. Edges in the graph represent dependencies between variables, encoded via either a conditional probability density or a deterministic function.

This graphical model serves as a mathematical counterpart to the conceptual model illustrated in Figure~\ref{fig:conceptualmodel}. The upper left-to-right path in Figure~\ref{fig:graphicalmodel} represents the time evolution of the physical asset state, represented by the random variables $S_t\sim p(s_t)$, while the lower path represents the time evolution of the digital state, represented by the random variables $D_t$. The graphical model encodes the tight two-way coupling between an asset and its digital twin. Information flows from the physical asset to the digital twin in the form of observational data, $o_t$, which are assimilated in order to update the digital state. Using the updated digital state, the models comprising the digital twin are used to predict quantities of interest, modeled at time $t$ as $Q_t\sim p(q_t)$. Information flows from the digital twin back to the physical twin in the form of control inputs, $u_t$, which are informed by the digital state and computed quantities of interest. These quantities all influence the reward for the timestep, $R_t\sim p(r_t)$.

Note that the graphical model depicted in Figure~\ref{fig:graphicalmodel} has both observations and control inputs occurring once per timestep, with observations occurring prior to control inputs. The structure of the graph (and the resulting algorithms we discuss throughout this work) can easily be adapted to situations in which this is not the case via a topological reorganization of the graph. For example, in the application presented in the following section, control inputs are issued first, with observational data acquired as a result.

The proposed graphical model (Figure~\ref{fig:graphicalmodel}) is sparsely connected in order to encode a set of known or assumed conditional independencies. In particular, the model encodes a Markov assumption for both physical and digital states, and the assumption that the physical state is only observable indirectly via data. Note that nodes in the graph can in general represent multivariate random variables, for example the digital state vector may consist of many parameters. In this case the graphical model does not specify independence structure between digital state parameters. The conditional independence structure defined by the graph allows us to factorize joint distributions over variables in the model. For example, we can factorize our belief about the digital state, $D_t$, quantities of interest, $Q_t$, and rewards, $R_t$, conditioned on the observed variables, namely the data, $O_t = o_t$, and enacted control inputs, $U_t = u_t$, for all timesteps up until the current time, $t = 0,...,t_c$, according to the structure of the proposed graphical model (Figure~\ref{fig:graphicalmodel}) as
\begin{linenomath}
\begin{align}\label{eqn:factorization_tc}
&p(D_0,...D_{t_c},Q_0,...,Q_{t_c},R_0,...,R_{t_c}  \mid  o_0,...,o_{t_c}, u_0,...,u_{t_c})\nonumber\\
&=\prod_{t=0}^{t_c} \left[\phi_t^{\textrm{update}}\phi_t^{\textrm{QoI}}\phi_t^{\textrm{evaluation}}\right],
\end{align}
where
\begin{align}
\phi_t^{\textrm{update}} &= p(D_t  \mid  D_{t-1}, U_{t-1}=u_{t-1}, O_t = o_t), \label{eqn:phiupdate}\\
\phi_t^{\textrm{QoI}} &= p(Q_t  \mid  D_t), \label{eqn:phiQoI}\\
\phi_t^{\textrm{evaluation}} &= p(R_t  \mid  D_t, Q_t, U_t = u_t, O_t = o_t).\label{eqn:phievaluation}
\end{align}
\end{linenomath}
Prediction is achieved by extending this belief state to include digital state, quantity of interest, and reward variables up until the chosen prediction horizon, $t_p$ (see the Methods section for details).

The factored representation \eqref{eqn:factorization_tc} serves two purposes. Firstly, the factors denoted $\phi$ and defined in \eqref{eqn:phiupdate}\textendash\eqref{eqn:phievaluation} are conditional probability distributions that reveal particular interactions that must be characterized by the computational models comprising the digital twin. Secondly, as we demonstrate in the following section, the factorization serves as a foundation for deriving efficient sequential Bayesian inference algorithms that leverage the modeled interactions in order to enable key digital twin capabilities such as asset monitoring, prediction, and optimization.

%
%
%
%
Note that the proposed probabilistic graphical model does not restrict the nature of the models comprising the digital twin. Each of these models could be physics-based (e.g., models based on discretized partial differential equations), data-driven (e.g., neural networks trained on historical or experimental data), or rule-based (e.g., state-transitions in a finite-state automaton). Note also that these models need not be specific to a single physical asset. Indeed, the models may be shared across a large fleet of similar assets, each with a digital state governing the asset-specific parameters used in these models. This enables efficient, centralized data and model management, and also enables each asset in the fleet to contribute to the continual improvement and enrichment of models by providing performance data relevant to a particular part of the state space.
\subsection*{Demonstration: Data-driven calibration and evolution of a UAV structural digital twin}
We instantiate the proposed approach for the structural digital twin of a UAV, with the target use case of managing and optimizing a large fleet of UAV assets (e.g., a fleet of autonomous package delivery vehicles). In particular, we demonstrate how the probabilistic graphical model can be used for the calibration, tailoring, and dynamic updating of computational models within the digital twin, so that they accurately reflect the current characteristics of a unique physical asset.

\begin{figure*}[b!]
\centering
\includegraphics[trim=0cm 0.5cm 0cm 0.5cm, width=0.99\linewidth]{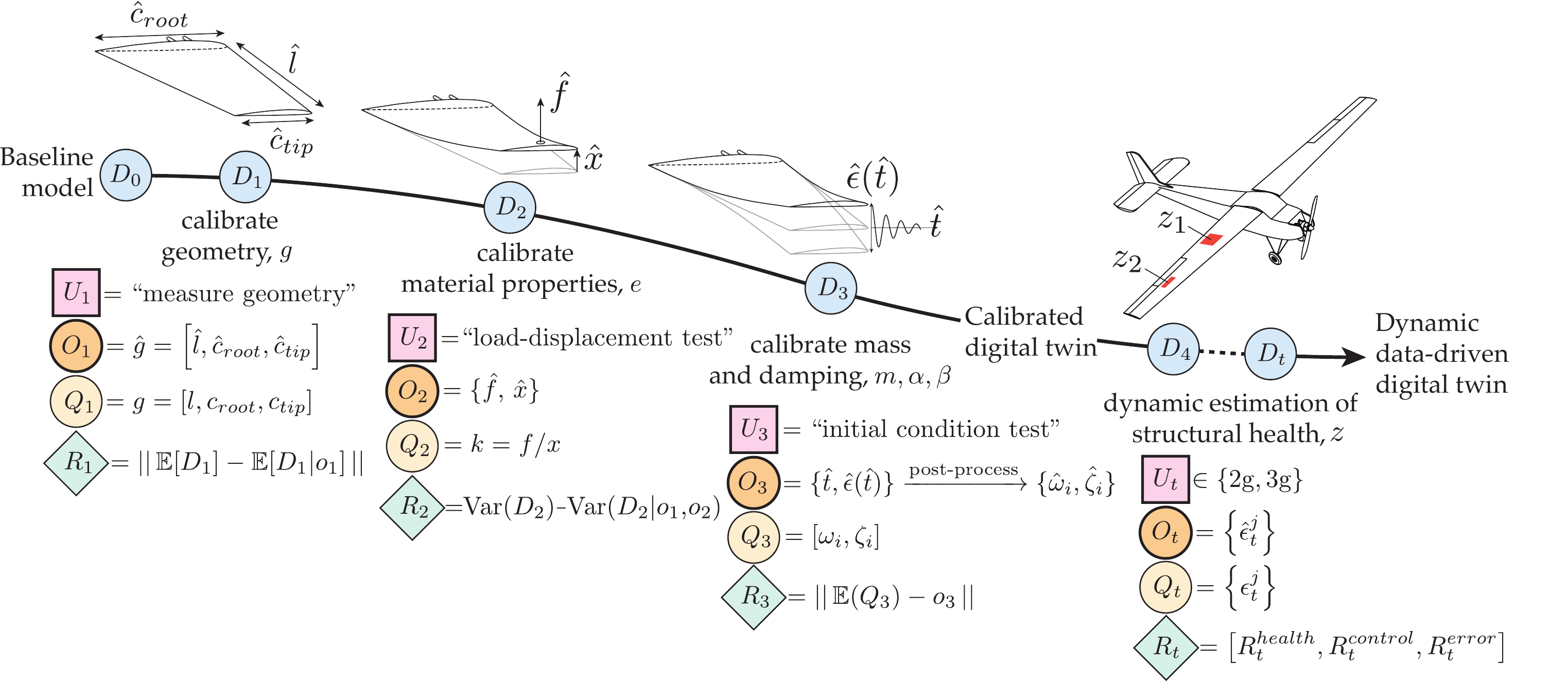}
\caption{Evolution of a UAV structural digital twin using the probabilistic graphical model formulation.}
\label{fig:experiment_flow}
\end{figure*}
We focus on two key stages in the lifecycle of any digital twin. First, we consider the initial calibration of the digital twin to an as-manufactured asset. While this type of experimental model calibration is commonplace throughout engineering, we here demonstrate how formulating the calibration process using our proposed probabilistic graphical model (Figure~\ref{fig:graphicalmodel}) ensures that the process is principled, scalable, and repeatable across the entire fleet of UAVs. Next, we show how the same probabilistic graphical model naturally extends in time beyond calibration and into the service life of each UAV. In our example, the calibrated UAV digital twin enables predictive structural simulation of the airframe. Dynamically evolving the digital twin based on incoming sensor data enables in-flight structural health monitoring, a capability which in turn enables dynamic health-aware mission planning, as well as simulation-based evaluation, certification, predictive maintenance, and optimized operations of the fleet.

In this example application we define the physical state, $S$, of the UAV asset to be its structural state. The physical state-space thus encompasses any conceivable structural variation between UAV assets, such as differences in geometry and material or structural properties. While many UAVs in the fleet share a common design, variation in material properties and manufacturing processes makes no two physical assets truly identical. Furthermore, once the asset enters operation, the physical state begins to evolve over time, for example due to maintenance events, damage, or gradual degradation. We seek to capture this variation by creating, calibrating, and dynamically evolving a structural digital twin for each unique UAV asset. We here present a summary of our formulation and results. Details on the physical asset, experimental setup, finite element structural models comprising the digital twin, and a full mathematical formulation are described in the Methods section.

For this application the digital state is defined to be:
\begin{linenomath}
\begin{equation}
d := \begin{bmatrix}
    g \\
    e \\
    m \\
    \alpha\\
    \beta\\
    z
\end{bmatrix}
\begin{array}{l}
    \MyLBrace{1ex}{\textrm{vector of geometric parameters}} \\
    \MyLBrace{1ex}{\textrm{Young's modulus scale factor}} \\
    \MyLBrace{1ex}{\textrm{vector of added point masses}} \\
    \MyLBrace{3ex}{\textrm{Rayleigh damping coefficients}} \\
    \MyLBrace{1ex}{\textrm{vector of structural health parameters}} \\
\end{array}
\label{eq:state}
\end{equation}
\end{linenomath}
This digital state comprises parameters of the UAV structural models that account for differences between UAV assets due to variability in materials, manufacturing, or operational history (geometry, $g$, Young's modulus scale factor, $e$, and other structural health parameters, $z$), parameters that represent unmodeled details of the UAV (the point masses $m$ account for physical UAV components that are not represented in the finite element model), and parameters that represent unmodeled physics (the Rayleigh damping coefficients are widely used in engineering to approximately model internal structural damping). By focusing on this set of parameters (while fixing all other model parameters at a nominal value) we obtain a digital twin that is tractable to calibrate and dynamically evolve, while sufficiently capturing the key differences in structural response between UAV assets. Figure~\ref{fig:experiment_flow} presents a summary of the probabilistic graphical model formulation for this application.
\paragraph{Results: Calibration Phase}
\begin{figure*}[b]
\centering
\includegraphics[trim=0cm 2.5cm 0cm 0cm, width=0.99\linewidth]{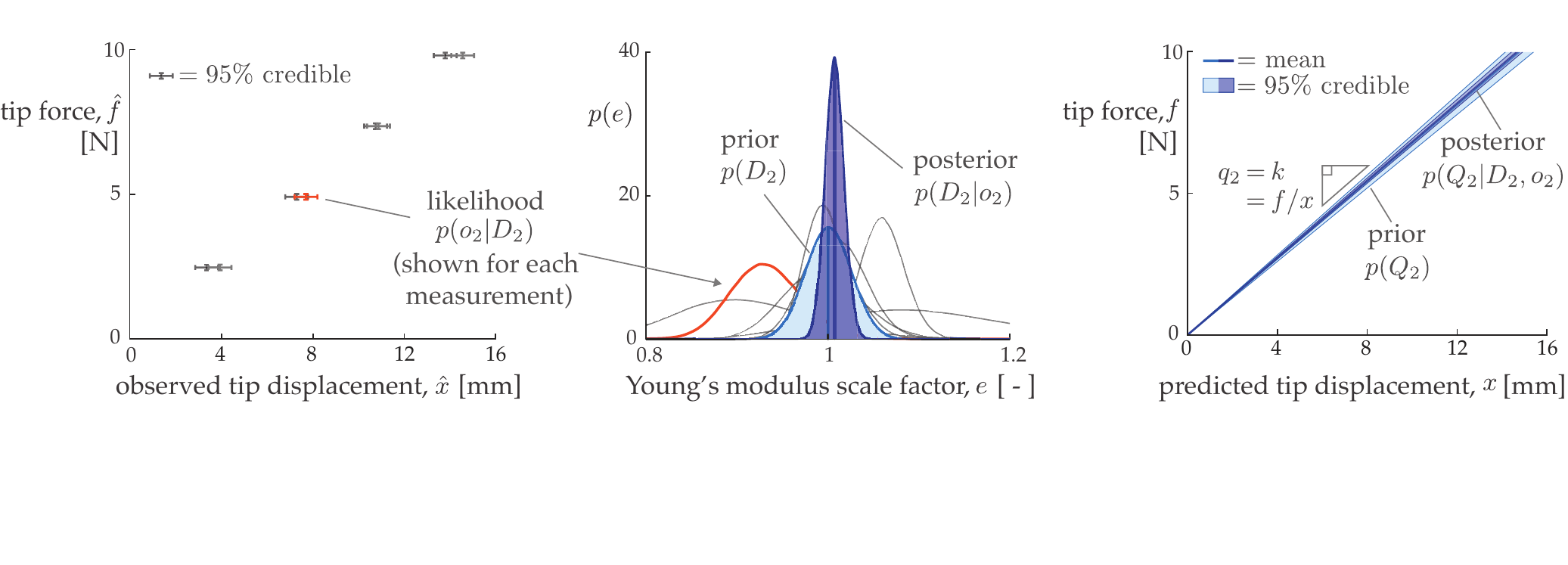}
\caption{Numerical results for the second calibration step ($t=2$). Left: Observational data, $o_2$. Center: Prior and posterior estimates of the component of the digital state, $D_2$, updated at this step. Right: Prior and posterior distributions of the quantity of interest, $Q_2$, for this step.}
\label{fig:stiffness_results}
\end{figure*}
We first consider the calibration phase in which the digital state parameters, and thus the computational models within the digital twin, are tailored so that they accurately reflect the unique characteristics of an as-manufactured physical asset. Each digital twin begins the calibration process with the same prior belief about the digital state parameters, $p(D_0)$. This distribution over model parameters defines a baseline UAV structural model based on design specifications, with uncertainty based on the degree of variability or confidence in each parameter. In the calibration phase, the timesteps, $t$, correspond to steps in the calibration process. At each step, $t=1,2,3$, we select an action $U_t = u_t$, which corresponds to conducting a specific experiment on the physical UAV asset. This experiment generates data, $o_t$, which inform an updated belief about the digital state, $p(D_t  \mid  O_t=o_t)$. Specifically, each experiment calibrates a subset of parameters in the digital state, and is designed to be conditioned only on previously calibrated parameters, with no dependence on parameters to be calibrated in a future step. Using this updated distribution of parameters in the computational model (described in the Methods section), the digital twin is used to compute quantities of interest, $p(Q_t  \mid  D_t, O_t=o_t)$, which describe the structural response of the UAV. We evaluate the success of the calibration procedure by estimating the reward, $p(R_t)$, at each stage. Here, values that are experimentally measured or derived from experimental data are denoted as hat variables, while the corresponding variable without a hat represents a computational estimate produced by digital twin models. We use braces ($\{\cdot\}$) to denote that there is an ensemble of data generated by repeated trials of an experiment.

At step $t=1$ we calibrate the geometric parameters, $g = [l, c_{root}, c_{tip}]$, where $l$ is the wing semi-span, $c_{root}$ is the chord length at the root, and $c_{tip}$ is the chord length at the tip. The control input, $u_1$, for this step is to physically measure each geometric parameter, producing measurements $\hat{g}$. Since these measurements are able to be taken accurately, the posterior uncertainty in the geometric parameters is negligibly small. Thus, the posterior distribution, $p(D_1  \mid  O_1=o_1, U_1=u_1)$, deterministically sets the geometric parameters to their measured values. The reward function for this calibration step measures the difference between prior and posterior estimates of each geometric parameter. The chosen norm could weight each parameter by its manufacturing tolerance, thus providing an overall measure of how well this particular UAV has been manufactured.

At step $t=2$ we calibrate the static load-displacement behavior of the structural model. In the digital state we update $e$, which is a scale factor applied to the Young's modulus (both longitudinal and transverse) of the carbon fiber material used in the wing skin. This scale factor allows us to adjust the computational model to account for material or manufacturing variation in the wing skin. The control input, $u_2$, is a decision to perform a static load-displacement test on the physical UAV asset. The observed data are a set of applied tip-load and measured tip-displacement pairs, $o_2 = \{\hat{f}, \hat{x}\}$, as shown in Figure~\ref{fig:stiffness_results} (left). Uncertainty in the applied load, $\hat{f}$, and measured tip-displacement measurement, $\hat{x}$, are both modeled by Gaussian distributions, with 95\% credible interval of widths of 20g and $1$mm respectively. Using this observed data, we perform a Bayesian update on our prior estimate in order to produce the posterior estimate $p(D_2  \mid D_1, O_2=o_2, U_2=u_2)$, as shown in Figure~\ref{fig:stiffness_results} (center). We see that the posterior mean value for this scale factor is 1.0073, indicating that the Young's modulus of carbon fiber in the digital twin model should be increased by 0.73\% to better match this particular physical UAV asset. The uncertainty in this parameter is also greatly reduced post-calibration. The quantity of interest for this step is a computational estimate of the coefficient of proportionality between tip load and tip displacement, as estimated by the digital twin models. The prior and posterior distributions of this quantity of interest are shown in Figure~\ref{fig:stiffness_results} (right). We evaluate the success of this calibration step by defining a reward that measures the reduction (achieved through calibration) in the variance of the random variable representing our knowledge of the scale factor $e$. As we will later see, this is of interest since variance in the Young's modulus scale factor $e$ represents model uncertainty that limits the ability of the digital twin to perform data assimilation during the operational phase.

In the final step $t=3$, we calibrate the dynamic response of the structural model. This depends on the mass distribution and damping properties of the wing, which are characterized by the digital state parameters, $m, \alpha,$ and $\beta$. In this example $m = [m_{servo}, m_{pitot}]$ defines points masses added to the model to represent the two identical servo motors and a single pitot tube fixed to the wing. The control input, $u_3$, for this step is a decision to perform an initial condition response experiment. This is done by applying an initial tip displacement and releasing the wing, recording data, $o_3$, in the form of strain, $\hat{\epsilon}$, as a function of time using a dynamic strain sensor. The strain data are post-processed to extract natural frequencies, $\hat{\omega}_i$ and damping ratios, $\hat{\zeta}_i$ for the first two bending modes $i=1,2$. As described in the Methods section, we formulate and solve an optimization problem to fit point masses, $m$, to the wing such that the natural frequencies of the first two bending modes predicted by the model, $\omega_1$, and $\omega_2$, match the experimental data, $\hat{\omega}_2$ and $\hat{\omega}_2$, as closely as possible while also matching the total wing mass. Using the calibrated natural frequencies we then compute the coefficients, $\alpha$, $\beta$, in the Rayleigh damping model. Together with the computed masses, this gives an estimate of the final calibrated posterior distribution $p(D_3  \mid  D_2, O_3=o_3, U_3=u_3)$. The quantities of interest for this step are the posterior computational estimates for modal parameters (frequencies and damping ratios). The reward for this step measures posterior predictive error, i.e., the difference between posterior estimates for the modal parameters and values that were obtained experimentally.

Prior and marginal posterior estimates for each component in the digital state are summarized in Table~\ref{fig:calibration_results_table}. We use $\mathcal{N}(\mu,\sigma)$ to denote a Normal distribution with mean $\mu$ and standard deviation $\sigma$. For sample distributions we report the sample mean followed by the sample standard deviation in parentheses.
\begin{table*}[t]
\centering
\includegraphics[trim=0cm 0cm 0cm 0.25cm, width=0.99\linewidth]{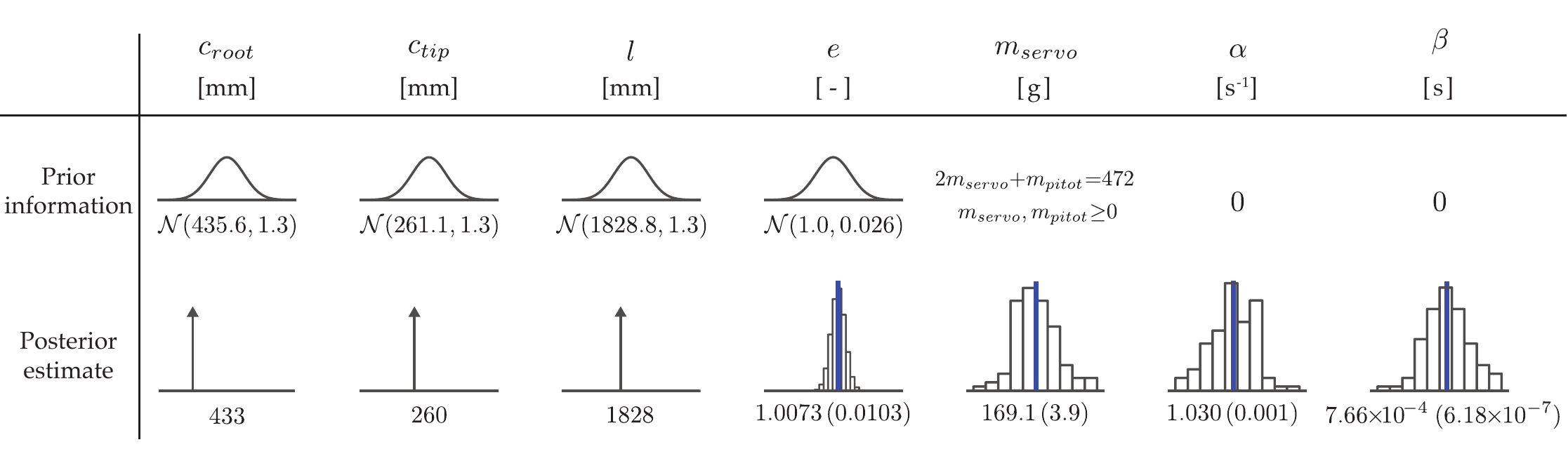}
\caption{Prior and posterior estimates for calibrated entries in the digital state. The first row shows prior information defining the initial estimate for each entry in the digital state. The second row shows posterior distributions for each entry in the digital state, which are the result of assimilating experimental data acquired via calibration experiments.}
\label{fig:calibration_results_table}
\end{table*}
\paragraph{Results: Operational Phase}
Next we extend the probabilistic graphical model to an operational phase in which the calibrated digital twin is deployed alongside the UAV. To demonstrate this application of the graphical model we simulate a demonstrative UAV mission consisting of successive level turns. This is an adaptation of the mission considered in a prior work \cite{kapteyn2020digitaltwin}. It illustrates how the proposed graphical model formulation permits us to naturally account for uncertainty from the calibration phase, as well as extend the digital twin capability to include planning, prediction, and evaluation.

In this phase, the timesteps $t=4,5,\ldots$ correspond to points in time during the UAV's mission. We fix the first five calibrated entries of the digital state (\ref{eq:state}) and dynamically estimate the remaining structural health parameters, $z$, as the health of the UAV evolves over time. In particular, variation in the structural health of the UAV is modeled by introducing two defect regions into the digital twin structural models. These defect regions are on the upper surface of the right wing, as shown in Figure~\ref{fig:experiment_flow}. We then define two structural health parameters within the digital state,
\begin{linenomath}
\begin{equation}
    z = [z_1, z_2] \in \{0, 20, 40, 60, 80\}\times\{0, 20, 40, 60, 80\},
\end{equation}
\end{linenomath}
where $z_1$ and $z_2$ define the percentage reduction in material stiffness applied to defect regions $1$ and $2$ respectively.

The digital twin is equipped with an internal model of how the structural health is expected to evolve, depending on which maneuvers are executed. This enables the digital twin to respond adaptively to its evolving structural health by determining optimal maneuvers for the UAV. In particular, at each timestep the digital twin issues a control input, $u_t \in \{2\textrm{g}, 3\textrm{g}\}$, which instructs the UAV to take the next turn at a bank angle corresponding to an aerodynamic load factor (the ratio of lift to weight) of either $2\textrm{g}$ or $3\textrm{g}$. Taking a turn at a steeper bank angle makes the path shorter, but also subjects the UAV to an increased aerodynamic load, which has a greater chance of worsening the UAV structural health.

The digital twin dynamically estimates the structural health of the UAV by using its calibrated internal models to assimilate observational data and adjust its predictions accordingly. For this demonstration the observational data at each timestep are noisy strain measurements,
\begin{linenomath}
\begin{equation}
  o_t = \{\hat{\epsilon}_t^j\}_{j=1}^{24},
\end{equation}
\end{linenomath}
from each of 24 uniaxial strain gauges, $j=1,\ldots,24$, on the upper surface of the wing (positioned near the defect regions as shown in Figure~\ref{fig:uavElements}).

During flight, the digital twin uses its estimate of the current structural health parameters in its internal structural models to provide a deeper analysis of the UAV structural integrity and consequent flight capability. In this example, the quantities of interest are defined to be computational estimates of the strain at strain gauge locations
\begin{linenomath}
\begin{equation}
  q_t = \{\epsilon_t^j\}_{j=1}^{24}.
\end{equation}
\end{linenomath}
This quantity is used to perform a posterior predictive check: the digital twin compares its posterior estimate of the strain with the observed strain in order to evaluate how well its models match reality. In practice this type of check can help validate other predictions made by the digital twin, such as modal quantities or the full stress and strain fields. Here, we quantify the posterior predictive error via the reward function
\begin{linenomath}
\begin{equation}
 r^{error}_t(o_t,q_t) =-\frac{1}{24}\sum_{j=1}^{24}\dfrac{|\hat{\epsilon}_t^j - \epsilon_t^j|}{\sigma^{sensor}}, \label{eq:reward1}
\end{equation}
\end{linenomath}
which measures the difference between observed strain measurements and strains predicted by the digital twin, normalized by an estimate of the sensor standard deviation, $\sigma^{sensor}$. Note that the negative sign ensures that any error is penalized via a negative reward. We define two additional reward functions targeted at different aspects of the mission. We define
\begin{linenomath}
\begin{equation}
  r^{health}_t(q_t) = \frac{\epsilon_{\text{max}} - \text{max}_j(\epsilon^j_t)}{\epsilon_{\text{max}}}, \label{eq:reward2}
\end{equation}
\end{linenomath}
as a measure of how far the UAV is from structural failure, as defined by a maximum allowable strain level, $\epsilon_{max}$. This term rewards the UAV for remaining in good structural health, as indicated by low predicted strain. Finally, we define
\begin{linenomath}
\begin{equation}
  r^{control}_t(u_t) = \left\{\begin{array}{rl}0.1 & \text{if } u_t=\text{3g}\\
  -0.1 & \text{if } u_t=\text{2g}\end{array}\right. \label{eq:reward3}
\end{equation}
\end{linenomath}
as a reward assigned to each applied control input. In this case the faster $3g$ turn is assigned a higher reward.

Prior to the mission, we solve a planning problem induced by the graphical model structure, as described in the Methods section. This involves using the digital twin models to predict how the structural health will evolve over the mission and computing a health-dependent control policy that optimizes a weighted sum of the expected rewards over the mission. For this demonstrative mission the computed control policy recommends that the UAV fly the more aggressive $\text{3g}$ maneuver until $z_1\geq60$, at which point it should fall back to the more conservative $\text{2g}$ maneuver. Note that in this case the control policy does not depend on $z_2$, since the digital twin's analysis reveals that this parameter has little influence on structural integrity, as measured by $R^{health}$.

\begin{figure*}[t]
\centering
\includegraphics[trim=0cm 0cm 0cm 0cm, width=\linewidth]{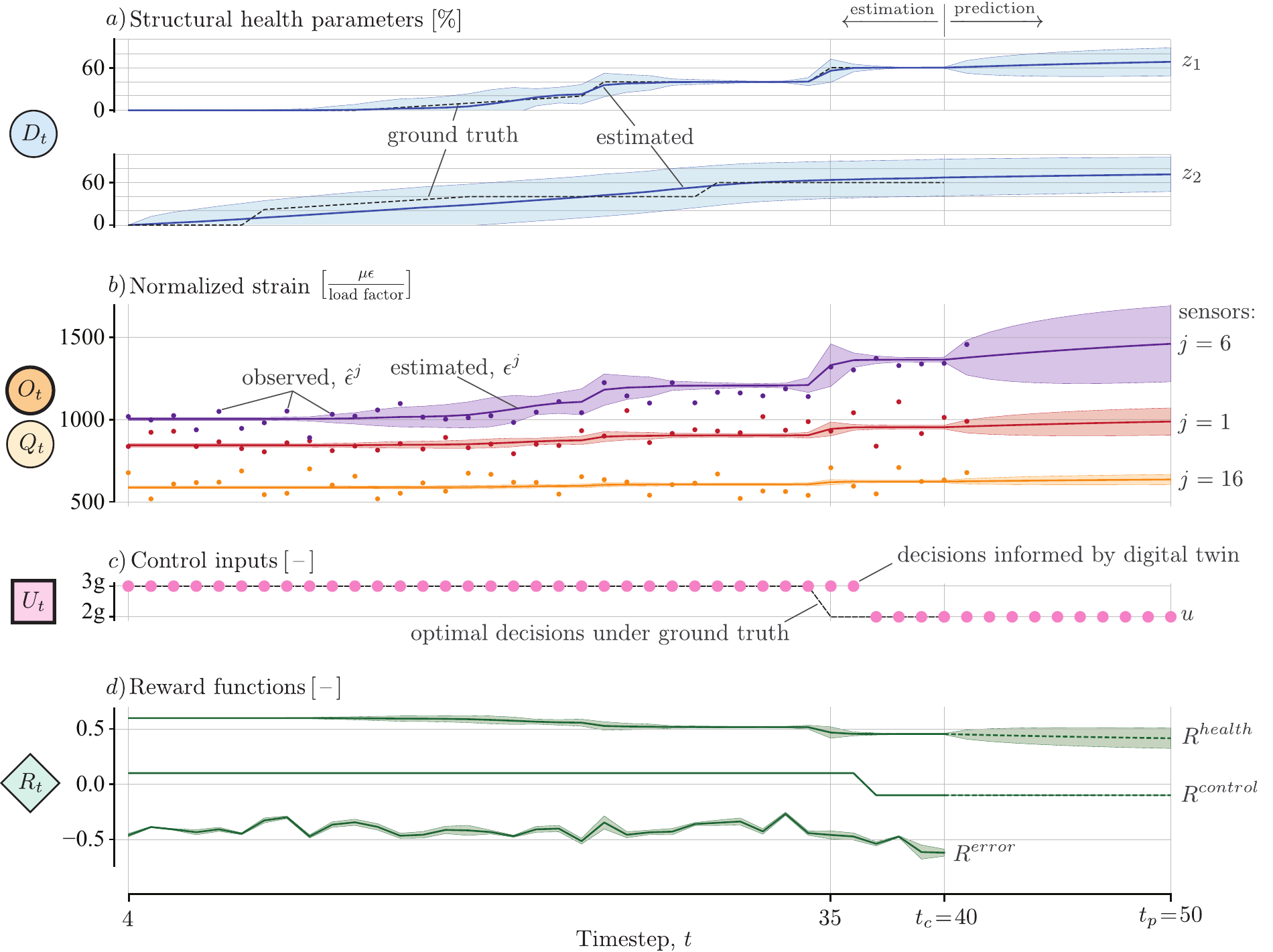}
\caption{Numerical results for the online phase of the simulated UAV mission. Probabilistic estimates are shown via the mean and two sigma uncertainty bands for the digital state (structural health parameters), quantity of interest (normalized strain), and reward functions ($R^{health}$, $R^{control}$, and $R^{error}$). The maximum a posteriori estimate is shown for the control input.}
\label{fig:online_results}
\end{figure*}

During the mission, the digital twin leverages its calibrated structural models to assimilate incoming sensor data, estimate how the structural health parameters and quantities of interest have evolved throughout the mission, and respond accordingly by suggesting control inputs, all with quantified uncertainty. Figure~\ref{fig:online_results} presents a snapshot of numerical results for the simulated mission at timestep $t_c=40$. The prediction horizon is set to $t_p=t_c+10$, so that the digital twin provides predictions up to $10$ timesteps into the future. In this demonstration we prescribe a ground truth evolution of the structural health parameters (as shown in Fig.~\ref{fig:online_results}a), but this is unknown to the digital twin. In fact, we deliberately prescribe the structural health evolution in a way that does not exactly match the digital twin's predictive model. Observed data are shown in Figure~\ref{fig:online_results}b for a subset of the sensors, while Figure~\ref{fig:online_results}a shows the digital twin estimates for each structural health parameter after assimilating this data. In this example, the digital twin is able to assimilate the observed data in order to maintain an accurate estimate of $z_1$ with relatively low uncertainty. On the other hand, the digital twin has relatively high uncertainty in its estimate of $z_2$, indicating that observed data are less informative about this parameter. Note that the earlier calibration phase plays a critical role in the data assimilation process, as it ensures that the digital twin models are accurate and reduces their uncertainty, thereby improving the performance of this online data assimilation.

Quantities of interest are shown in Figure~\ref{fig:online_results}b for a subset of sensor locations. Note that uncertainty in the quantities of interest is due to a combination of uncertainty in the structural health parameters, $z$, and uncertainty in the Young's modulus scale factor, $e$, carried forward from the calibration phase via our unified graphical model. Control inputs estimated by the digital twin are shown in Figure~\ref{fig:online_results}c. We see that the digital twin is able to respond intelligently to the degrading structural health by switching to less aggressive maneuvers within two timesteps of when the (unknown) ground truth UAV structural health necessitated this change. The reward functions, shown in Figure~\ref{fig:online_results}d, reflect the structural health and control favorability of the UAV, as well as the predictive accuracy and uncertainty of the digital twin models.

%% file: sections/3_discussion.tex
\section*{Discussion}\label{sec:conclusion}
The probabilistic graphical model formulation advances the field of digital twins by clearly defining the elements comprising an abstract representation of an asset-twin system, identifying the interactions between elements that need to be modeled, and incorporating end-to-end uncertainty quantification via a Bayesian inference framework. Our model demonstrates how the digital twin paradigm incorporates dynamic updating and evaluation of computational models into a data assimilation and feedback control loop. These computational models offer valuable insights, unattainable through raw observational data alone, which can be leveraged for improved data-driven modeling and decision making \cite{brunton2019data}.

This paper presented an example application of the model, focused on a structural digital twin for a UAV. We first formulated a series of calibration experiments, where measured data are used to tailor the digital twin models to the as-manufactured specifications of the UAV. We then demonstrated how the calibrated digital twin could be leveraged in the operational phase of the UAV for dynamic in-flight health monitoring and adaptive mission planning. This application was chosen in order to demonstrate the process of defining quantities and instantiating the proposed graphical model. It also highlights how the flexibility of the model allows it to extend across phases of the asset lifecyle. In our demonstrative application, the graphical model enables principled data assimilation for digital twin adaptation, leveraging of the digital twin models for analysis and decision-making, and end-to-end uncertainty quantification. Beyond this UAV application, the declarative and general nature of the graphical model mean that it can be applied to a wide range of applications across engineering and science. Indeed, this application is just one example of a much broader class of next-generation intelligent physical systems that need to be monitored and controlled throughout their lifecycle.

Many opportunities exist to integrate specialized or application specific formulations for specific tasks within our probabilistic graphical model framework. For example, our structural health monitoring formulation could be extended to enable detection of unknown or anomalous states by integrating the graphical-model-based formulation for damage detection and classification presented in~\cite{dzunic2017bayesian}.

Limitations of the proposed framework include the challenge of defining and parameterizing the models comprising the digital twin. A central aspect of this challenge is a need to quantify and manage model inadequacy (sometimes referred to as model discrepancy \cite{brynjarsdottir2014}), a topic that is beginning to receive more attention throughout computational science. Another open challenge is sensor design for digital twins, as highlighted in our UAV structural health monitoring demonstration. In our graphical model this could be formalized through the control theoretic notion of observability, i.e., whether the available observational data combined with carefully selected control inputs is adequate to inform the digital state.
Finally, computational resource limitations remain a challenge for the realization of predictive digital twins. The probabilistic inference and planning procedures developed in this paper, even when accelerated via approximation techniques, require many evaluations of the high-fidelity physics-based models comprising the digital twin. Such models require approximations such as reduced-order modeling, surrogate modeling, and other compression techniques~\cite{hartmann2018model}.

%% file: sections/4_methods.tex
\section*{Methods}\label{sec:methods}
\subsection*{Physical UAV asset}
The physical asset used for this work is a fixed-wing UAV testbed vehicle~\cite{salinger2020hardware}. A rendering of the vehicle is shown in Figure~\ref{fig:uavElements}. The fuselage, empennage, and landing gear are from an off-the-shelf Giant Telemaster airplane. The wings have been custom designed with plywood ribs and carbon fiber skin, in order to mimic the structure of a commercial or military-grade vehicle, albeit at a reduced scale. The wings are connected to the fuselage via two carbon rods that extend 25\% of the way into each wing. The electric motor, avionics, and sensor suite are a custom installation. Further details on the sensors used for this work are provided in the experimental methodology section.

The calibration experiments in this work were conducted in a laboratory environment. As the calibration is focused only on the aircraft wings, they are detached from the fuselage and mounted to a fixed support. The wings are mounted upside-down so that typical aerodynamic forces can be more easily applied as downward forces.
\subsection*{Computational models comprising the UAV digital twin}
The computational model at the heart of the UAV structural digital twin is a physics-based model for the linear elastic structural response of the aircraft~\cite{kapteyn2020digitaltwin}. This model relates a time-varying load on the aircraft (e.g., wing loads for a given aircraft maneuver), to its structural displacement. We adopt a finite element spatial discretization, so that this relationship can be expressed as the second-order ordinary differential equation
\begin{linenomath}
\begin{equation} \label{eqn:ODE}
\text{M}(d)\ddot{\mathbf{x}}(t) + \text{V}(d)\dot{\mathbf{x}}(t) + \text{K}(d)\mathbf{x}(t) = \mathbf{f}(t).
\end{equation}
\end{linenomath}
Bolded quantities represent vectors of nodal quantities. In this case, $\mathbf{x}$, $\dot{\mathbf{x}}$, and $\ddot{\mathbf{x}}$ denote the displacement, velocity, and acceleration,  respectively, while $\mathbf{f}$ denotes the applied force at each node in the finite element mesh. The mass, damping, and stiffness matrices, $\text{M}(d)$, $\text{V}(d)$, and $\text{K}(d)$ respectively, are parametrized by the digital state, $d$. As defined by \eqref{eq:state}, the digital state is an asset-specific parameter vector that reflects the unique geometric and structural properties of each individual UAV.

The static load-displacement experiment used to generate data at step $t=2$ in the calibration application is simulated by the digital twin via the static version of \eqref{eqn:ODE},
\begin{linenomath}
\begin{equation}\label{eqn:static}
\text{K}(d)\mathbf{x} = \mathbf{f}.
\end{equation}
\end{linenomath}
Note that in the load-displacement calibration experiment, we set the applied loads vector, $\mathbf{f}$, to represent a point load of magnitude, $f$, near the tip. We solve the model to compute the displacement vector, $\mathbf{x}$, and post-process the solution to extract the tip displacement, $x$. Through a computational study using the static elasticity model \eqref{eqn:static}, it is found that the aggregate wing stiffness, $k$, i.e., the coefficient of proportionality between an applied tip load, $f$, and the tip displacement, $x$, depends linearly on the Young's modulus scale factor applied to the model, $e$. In particular, we establish the relationship
\begin{linenomath}
\begin{equation}\label{eqn:ktoe}
    k = 0.5752e +  0.1018.
\end{equation}
\end{linenomath}
While the calibration process focuses on a tip load, note that once calibrated the digital twin is able to accurately simulate the deformation of the wing to \emph{any} load, an example of the power of computational models for prediction and analysis. This is demonstrated in the operational phase, where we use \eqref{eqn:static} to predict the strain field in the wing while the aircraft is performing a steady level turn. For notational convenience, we define a function
\begin{linenomath}
\begin{equation}
  \epsilon^j(d,u)
  \label{eqn:strain}
\end{equation}
\end{linenomath}
which encapsulates the result of the following algorithm: 1) compute the wing loading, $\mathbf{f}$, for a given maneuver, $u$, using an ASWING \cite{drela1999integrated} aerostructural model of the aircraft, 2) apply the computed load to \eqref{eqn:static} to compute the displacement field, $\mathbf{x}$, and 3) post-process this displacement field to compute the strain field and extract the strain value corresponding to sensor $j$.

The dynamic response of the aircraft structure can also be characterized by the digital twin via an eigenanalysis. The computational model \eqref{eqn:ODE} can be adapted to perform an eigenanalysis of the form
\begin{linenomath}
\begin{equation}\label{eqn:eigen}
\text{K}(d)\tilde{\mathbf{x}}_i = \text{M}(d)\omega_i^2 \tilde{\mathbf{x}}_i
\end{equation}
\end{linenomath}
where $\tilde{\mathbf{x}}_i$ is a nodal displacement vector representation of the mode shape, and $\omega_i$ is the natural frequency for mode $i$. This model is used at step $t=3$ of the calibration procedure.

We adopt a Rayleigh damping model~\cite{chopra2012dynamics} which defines the damping matrix, $\text{V}$, as a combination of mass-proportional damping and stiffness proportional damping as follows:
\begin{linenomath}
\begin{equation}
  \text{V}(d) = \alpha \text{M}(d)+ \beta \text{K}(d).
\end{equation}
\end{linenomath}
Under this model the $i$th modal damping ratio, $\zeta_i$, is given by
\begin{linenomath}
\begin{equation}\label{eqn:zeta}
    \zeta_i = \frac{\alpha}{2}\frac{1}{\omega_i} + \frac{\beta}{2}\omega_i,
\end{equation}
\end{linenomath}
where $\omega_i$ is the $i$th natural frequency.

During calibration we focus on the first two bending modes, $i=1,2$, as these are clearly evident in our experimental data. However, the calibrated computational model is not limited to these modes---indeed the digital twin enables us to compute higher bending modes as well as torsional and skin buckling modes, which are difficult to characterize experimentally.

In this demonstration, the structural health parameters, $z_1$, and $z_2$, represent a percentage reduction in stiffness in specific regions of the aircraft wing. This is implemented in the finite element models, \eqref{eqn:ODE}, \eqref{eqn:static}. and \eqref{eqn:eigen}, by identifying elements within these regions and reducing their Young's modulus parameters. Note that in practice these structural health parameters could correspond to specific causes of damage, for example the length of a crack or the extend of a delamination. In our demonstration, these parameters are designed to simply account for the \emph{effect} of any damage or degradation present in the corresponding region. Further discussion on this choice of structural health parameterization can be found in \cite{kapteyn2020digitaltwin}.
\subsubsection*{Prior estimate for the digital state}
Our prior estimate of the geometric parameters, $g = [l, c_{root}, c_{tip}]$, is based on the nominal or as-designed value for each parameter, as stated on technical drawings for the wing design that were provided to the wing manufacturer. The allowable manufacturing tolerance is stated in these drawings as $\pm2.5$mm. Thus, our prior estimate of each geometric parameter is modeled as a Gaussian distribution, where the mean corresponds to the nominal design value, and the standard deviation is set to give a 95\% credible interval width equal to the manufacturing tolerance.

We set the prior distribution over the carbon fiber Young's modulus scale factor, $e$, to be a Gaussian distribution with mean $1$ and a $95\%$ credible interval equal to $\pm 5\%$ variability. This prior uncertainty is an expert-driven estimate that could be refined over time as more wings are manufactured and the degree of variability in material properties is characterized based on manufacturing data.

We add three point masses to the model: two have mass $m_{servo}$ and represent the servomotor hardware that actuates each aileron, while the third has mass $m_{pitot}$ and represents the pitot tube attached to the wing tip. The location of each of these hardware components is measured on the physical asset and fixed in the digital twin model. It is not possible to measure the mass of each component individually as they are fixed to the wing during manufacturing. Instead, we measure the total weight of the wing hardware and compare this with the mass accounted for in the computational model. A discrepancy of $472$g is identified. Thus, our prior information about these point masses comes in the form of the constraints:
\begin{linenomath}
\begin{align}
    2m_{servo} + m_{pitot} &= 472,\label{eqn:mconstraint1}\\
    m_{servo}, m_{pitot} &\geq 0\label{eqn:mconstraint2}.
\end{align}
\end{linenomath}

Finally, in this work we consider the prior estimates on the Rayleigh damping coefficients, $\alpha$ and $\beta$, to both be zero. This can be interpreted as having no damping in the model before the damping is experimentally calibrated. We also assume that the structural health parameters, $z$, are known to be zero throughout the calibration phase, i.e., we assume that no damage or degradation occurs until the asset enters operation.

\subsection*{Methodology for experimental calibration of the UAV Digital Twin}
This section provides additional details on the experimental and computational procedures used for each step of the calibration procedure.

\paragraph{Step 1: Calibrate geometry, $g$\,:}
The first calibration step is focused on updating the components of the digital state estimate corresponding to the geometric parameters $g = [l, c_{root}, c_{tip}]$.

To update the prior estimate over these parameters into the posterior, $p(D_1  \mid  O_1=o_1)$, we measure the as-manufactured wing geometry. The measured values constitute the observational data, $o_1 = [\hat{l}, \hat{c}_{root}, \hat{c}_{tip}]$. The maximum measurement error is estimated to be $\pm 0.5$mm. While each measurement could be represented by a Gaussian distribution and incorporated via a Bayesian update, the resulting posterior uncertainty would be practically negligible. We instead ignore this uncertainty, setting the posteriors to delta distributions centered on the measured values. This provides computational savings throughout the remainder of the calibration and allows us to use a fixed computational mesh in the digital twin.

\paragraph{Step 2: Calibrate material properties, $e$\,:}
We calibrate the Young's modulus using a static load-displacement experiment. A known mass is placed on the main spar of the wing, 5cm from the wing tip, generating a static tip load with magnitude $\hat{f}$. To account for error in the applied force, as well as the position of the force, we model the uncertainty using a Gaussian distribution with a 95\% credible interval equivalent to $\pm10$g. The resulting static tip displacement, $\hat{x}$, is then measured, with measurement error modeled as an independent Gaussian with 95\% credible interval equal to $\pm1$mm. This process is repeated for eight total measurements as shown in Figure~\ref{fig:stiffness_results} (left): two times each for applied masses of 250g, 500g, 750g, and 1000g. Thus, the observational data is a set of eight measured load-displacement pairs, $o_2 = \{\hat{f},\hat{x}\}$. Each measured load-displacement pair is converted into an estimate of the aggregate wing stiffness, $\hat{k} = \hat{f} / \hat{x}$, then each of these estimated stiffness values is converted into an equivalent estimate of the Young's modulus scale factor, $\hat{e}$, via \eqref{eqn:ktoe}.

The observational data, $o_2$, is incorporated into the estimate of $e$ according to the Bayesian update formula
\begin{linenomath}
\begin{equation}
    p(e  \mid  \hat{e}) \propto p(\hat{e} \mid e) \ p(e)
\end{equation}
\end{linenomath}
where $p(e)$ is the Gaussian prior distribution and $p(\hat{e} | e)$ is the likelihood function corresponding to a single measurement, $\hat{e}$, which in this case is non-Gaussian (due to the $1/\hat{x}$ dependence). We estimate this likelihood density by sampling and fitting a kernel density. In particular, we draw $10^6$ sample pairs from the Gaussian distributions centered on $\hat{f}$ and $\hat{x}$ respectively. Each sample pair is transformed into a sample of $\hat{e}$ using the procedure described above. We then fit a kernel density estimate to these samples, which serves as an estimate of the likelihood $p(\hat{e}  |  e)$. Each of these likelihood functions are shown in Figure~\ref{fig:stiffness_results} (center).

The Bayesian update is performed by incorporating each measurement iteratively using a standard particle filter approach. For each measurement, $\hat{e}$, we first draw $10^6$ samples from the prior distribution, $p(e)$, and assign each sample a uniform weight. We then scale the weight of each sample by the likelihood, $p(\hat{e}  |  e)$, before re-normalizing the weights. This results in a weighted-particle approximation of the posterior density after assimilating a single measurement. This process is repeated for each of the eight measurements taken, incrementally updating the posterior estimate. After all eight measurements are incorporated, we arrive at the posterior estimate, $p(e  |  O_2=o_2)$.  From this weighted-particle approximation of the posterior density we can estimate the mean and standard deviation (reported in Table \ref{fig:calibration_results_table}), as well as a posterior 95\% credible interval for $e$. This credible interval can be translated into a credible interval in terms of the quantity of interest, $k$, which is shown in Figure~\ref{fig:stiffness_results} (right). Finally, we can draw samples from the posterior using inverse transform sampling applied to the empirical cumulative density function. This enables us to propagate forward the uncertainty in $e$ into the next calibration step.

\paragraph{Step 3: Calibrate mass and damping, $m, \alpha,\beta$\,:}
In the final calibration step, we seek to calibrate the dynamic response of the digital twin model by adding point masses that represent unmodeled hardware, as well as determining the appropriate coefficients for a Rayleigh damping model.

We first describe the experimental data, $o_3$, used for this step. We experimentally characterize the dynamic response of the wing using an initial condition response experiment. We apply an initial tip displacement of $10$mm and release, resulting in a decaying oscillation of the tip displacement. During this oscillation we collect dynamic strain data, $\hat{\epsilon}(\hat{t})$, at time points, $\hat{t}$, measured from the moment the wing is released. Dynamic strain is measured using thin flexible piezoelectric patches that can be applied to different regions of the wing. These sensors communicate wirelessly, simplifying integration with the wing system. Data are taken at a sampling frequency of 100kHz, downsampled to 2kHz during post-processing. This experiment is repeated five times. From each raw dataset, we generate a power spectrum via fast Fourier transform. The power spectra for all experiments show two clear peaks corresponding to the first and second bending modes of the wing. We extract the locations of these peaks, which correspond to the first and second damped natural frequencies, $\hat{\omega}_i^d$. Using these frequencies, we seek to construct a two-mode reconstruction of the measured signal, $\hat{\epsilon}$, of the form
\begin{linenomath}
\begin{align}
    \hat{\epsilon}^{reconstructed} =& a_1 e^{-b_1 t}\cos(2\pi\hat{\omega}_1^d(\hat{t}-c_1))\nonumber\\
    +& a_2 e^{-b_2 t}\cos(2\pi\hat{\omega}_2^d(\hat{t}-c_2)).
\end{align}
\end{linenomath}
We fit this model to the measured data by solving a non-linear least-squares problem for the coefficients, $a_1,a_2,b_1,b_2,c_1,c_2$, using a Levenberg-Marquardt algorithm. From this model we can extract the experimentally derived damping ratios, $\zeta_i$, as well as the undamped natural frequencies, $\hat{\omega}_i$, by solving the following system of equations:
\begin{linenomath}
\begin{align}
    \hat{\omega}_i &= \frac{\hat{\omega}^d_i}{\sqrt{1-\hat{\zeta}_i^2}},\\
    \hat{\zeta}_i &= \frac{b_i}{2\pi\hat{\omega}_i}.
\end{align}
\end{linenomath}
This procedure is repeated for each dataset, resulting in a total of five experimental estimates for the modal frequencies and damping ratios, $o_3 = \{\hat{\omega}_i, \hat{\zeta}_i\}$, for the first two bending modes, $i=1,2$. We average across all experimental datasets. In the following we refer to these averages by $\hat{\omega}_i$ and $\hat{\zeta}_i$ for simplicity.

Recall that our goal is to update the belief at step $t=2$, namely $p(D_2  |  D_1, O_2=o_2, U_2=u_2)$, into a belief state at $t=3$, namely $p(D_3  |  D_2 ,O_3=o_3, U_3=u_3)$. We begin by drawing a sample of the calibrated parameters from the previous belief state. Since the calibrated geometric parameters are deterministic, this amounts to sampling a value for the remaining parameter, $e$, from the posterior distribution $p(e  |  O_2=o_2)$ computed during the previous calibration step.

Given the observational data $o_3$, and a sampled value for $e$, we seek to add to our sample a computed value for the point masses $m$. Modifying these point masses in the computational model changes the mass matrix, $\textrm{M}(d)$. Through the computational model \eqref{eqn:eigen}, this in turn changes the natural frequencies, $\omega_i$, predicted by the model. Using this model we formulate an optimization problem to fit the point masses, $m = [m_{servo}, m_{pitot}]$, so that the natural frequencies of the first two bending modes predicted by the model, $\omega_1$, and $\omega_2$, match the experimental data, $\hat{\omega}_2$ and $\hat{\omega}_2$, as closely as possible, while obeying the mass constraints \eqref{eqn:mconstraint1}--\eqref{eqn:mconstraint2}. The objective function used in this optimization is the sum of relative frequency errors, and the optimization problem is solved using a Nelder-Mead gradient-free optimization method.

Finally, using the calibrated natural frequencies we compute the coefficients, $\alpha$, $\beta$, in the Rayleigh damping model according to the system of equations
\begin{linenomath}
\begin{equation}
\frac{1}{2}\begin{bmatrix}
1/\omega_1 & \omega_1 \\
1/\omega_2 & \omega_2
\end{bmatrix}\begin{bmatrix}
\alpha\\
\beta
\end{bmatrix}
=
\begin{bmatrix}
\hat{\zeta}_1 \\
\hat{\zeta}_2
\end{bmatrix}.
\end{equation}
\end{linenomath}
Note that this system is derived from \eqref{eqn:zeta}, and ensures by construction that the computational damping ratios, $\zeta_i$, exactly match the experimentally estimated damping ratios, $\hat{\zeta}_i$, for both modes $i=1,2$. The computed values for $m,\alpha,$ and $\beta$, combined with the sampled value for $e$ and the calibrated geometric parameters $g$, constitute a sample from the updated belief state. We repeat this entire optimization procedure for 100 samples in order to build up a sample-based approximation of the final calibrated estimate of the digital state.

The quantities of interest for this step are the posterior computational estimates of the modal frequencies and damping ratios, $\omega_i$, and $\zeta_i$. The reward is the average discrepancy between these values and the corresponding experimentally estimated values, $\hat{\omega}_i$ and $\hat{\zeta}_i$ respectively.
\subsection*{Methodology for dynamic estimation of UAV structural health}
\paragraph*{Mission Simulation}
We simulate a demonstrative UAV mission consisting of successive level turns. We suppose that throughout the mission the UAV undergoes structural degradation. For this demonstration we prescribe a ground truth evolution of the two structural health parameters, $z$, as shown in Figure \ref{fig:online_results}. This ground truth trajectory is intended to represent periods of gradual degradation in structural health, as well as sudden changes in structural health, e.g., due to discrete damage events. The ground truth state is unknown to the digital twin, and is only used to simulate sensor data.

We discretize time such that each timestep $t\geq4$ occurs during a quasi-steady section of a turn. This allows us to simplify the problem and ignore the transient period between turns. We simulate noisy strain measurements by using the digital twin structural models, \eqref{eqn:strain}, to compute the true underlying strain while assuming a truth value of $e=1.0073$ (corresponding to the posterior mean). We then corrupt the true strain for each sensor with zero mean uncorrelated Gaussian noise with standard deviation equal to $150$ microstrain.

The simulation is implemented and dynamically executed using ROS \cite{quigley2009ros}. One ROS node represents the simulated UAV asset, and another represents the digital twin. At each timestep, the UAV simulator generates strain data based on its current state and most recent control input, and passes this to the digital twin. The digital twin maintains an instance of the probabilistic graphical model, which it uses to perform planning, data assimilation, and estimation, as described in the following sections. The digital twin completes the timestep by deciding on the subsequent control input and passing this to the UAV.
\paragraph*{Bayesian Inference Formulation}\label{sec:monitoring}
This section describes how the proposed graphical model enables us to formulate asset monitoring as a sequential Bayesian inference task. In this application, our goal at each timestep is to estimate the entire mission history of digital states, quantities of interest, and rewards. Thus, our target belief state is the \textit{smoothing} distribution given by \eqref{eqn:factorization_tc}. As described in the Results section, this distribution can be factorized according to the structure of the proposed graphical model to reveal the interaction factors, \eqref{eqn:phiupdate}\textendash\eqref{eqn:phievaluation}. We first describe how we compute each of these factors, before discussing how they are combined in a Bayesian inference algorithm.

The digital twin update factor, $\phi_t^{\textrm{update}}$ (Eq.~(\ref{eqn:phiupdate})), quantifies how the digital state is updated at each timestep, conditioned on the digital state and control input at the previous timestep, and any newly acquired observational data. Note, in this demonstration we do not update our estimate of the Young's modulus scale factor, $e$, beyond the calibration phase. However, we do account for the posterior uncertainty in $e$ when updating our estimate of the structural health parameters, $Z_t \sim p(z_t)$, at each timestep $t$. The update factor can thus be written in the form
\begin{linenomath}
\begin{align}
\phi_t^{\textrm{update}} =& p(Z_t  \mid  Z_{t-1}, U_{t-1}=u_{t-1}, O_t=o_t)\\
 =& \int p(Z_t  \mid  Z_{t-1}, e, U_{t-1}=u_{t-1}, O_t=o_t)p(e) \, \textrm{d}e
\end{align}
\end{linenomath}
The digital twin update term can be factorized further using Bayes' rule and conditional independence to explicitly separate the contributions of data assimilation and state dynamics to the digital twin update:
\begin{linenomath}
\begin{equation}\label{eqn:predictorcorrector}
\phi_t^{\textrm{update}} \propto \phi_t^{\textrm{dynamics}}\phi_t^{\textrm{assimilation}},
\end{equation}
\end{linenomath}
where
\begin{linenomath}
\begin{align}
&\phi_t^{\textrm{dynamics}} = p(Z_t  \mid  Z_{t-1}, U_{t-1}=u_{t-1})\\
&\phi_t^{\textrm{assimilation}} = p(O_t=o_t \mid  Z_t) = \int p(O_t=o_t  \mid  Z_t, e)p(e) \, \textrm{d}e
\end{align}
\end{linenomath}
The form~\eqref{eqn:predictorcorrector} gives a predictor-corrector type update policy, commonly seen in hidden Markov models, Kalman filtering, particle filtering, etc. The first term in the right-hand side of~\eqref{eqn:predictorcorrector} corresponds to a prediction forward in time based on the control-dependent transition dynamics of the system. To define the state transition dynamics, $\phi_t^{\textrm{dynamics}}$, we assume that the probability of damage progression in each defect region is known, fixed, and conditionally independent given the load on the aircraft wing (given by the load factor $u_t$). In particular, we suppose that the structural health in each defect region has a $0.05$ probability to worsen by $20\%$ under a $\text{2g}$ maneuver, and a $0.10$ probability to worsen by $20\%$ under a $\text{3g}$ maneuver. Note that this model is not a perfect reflection of the true evolution of the UAV structural health that we prescribe in this simulated mission. This is intentional, and highlights the fact that the digital twin can leverage data to perform useful asset monitoring, even with an imperfect model of the underlying asset dynamics. Also note that in practice this transition probability model could be more complex, and would typically be estimated from offline experiments, physics-based damage progression simulations, or historical data from past flights of similar UAVs.

The second term in the right-hand side of~\eqref{eqn:predictorcorrector} encapsulates a correction or reduction in uncertainty via data assimilation. We equip the digital twin with a sensor model, namely that the strain measurements for sensors $j=1,\ldots,24$ are given by
\begin{linenomath}
\begin{equation}
  \hat{\epsilon}^j_t = \epsilon^j_t + v_t,
  \label{eqn:sensormodel}
\end{equation}
\end{linenomath}
where $v_t \sim N(0,\sigma^{sensor})$ represents zero mean, uncorrelated Gaussian noise with a standard deviation of $\sigma^{sensor}=125$ microstrain. Note that this sensor model does not match our simulated measurements exactly.

In (\ref{eqn:sensormodel}), $\epsilon^j_t$ is the digital twin estimate of the true strain, which is computed according to \eqref{eqn:strain}. The predicted strain is a function of the digital state, in particular the Young's modulus scale factor, $e$, and the structural health parameters, $z_t$. We do this by first sampling a set of $N=30$ values from the calibrated estimate for $e$ (shown in Figure~\ref{fig:calibration_results_table}). For each sample $e^k$, we compute the predicted strain, $\epsilon_t^j$, for each possible damage state, $z_t$. We then compute $p(\hat{\epsilon}_t^j  \mid  z_t, e^k)$ via the measurement model, \eqref{eqn:sensormodel}, and average across samples to compute the assimilation factor used to update our estimate of $Z_t$,
\begin{linenomath}
\begin{equation}
  p(O_t=o_t \mid  Z_t) \approx \prod_{j=1}^{24}\frac{1}{N}\sum_{k=1}^N p(\hat{\epsilon}_t^j  \mid  Z_t, e^k).
\end{equation}
\end{linenomath}

The factor $\phi_t^{\textrm{QoI}}$ (Eq.~(\ref{eqn:phiQoI})) encapsulates the process of executing the updated models comprising the digital twin in order to estimate quantities of interest that characterize the physical asset. In this example our quantities of interest are the true strain, $\epsilon_t^j$, at each sensor location, $j$, which are estimated by propagating the uncertainty in $e$ forward through the strain model \eqref{eqn:strain} for each possible damage state, $z_t$. Here we again use $N=30$ samples drawn from the calibrated estimate of $e$. During inference this factor is multiplied by our posterior estimate of $Z_t$ to give the posterior estimate of $Q_t$. Note that in this example the quantity of interest has an implicit dependence on the control inputs, since they affect the load on the airframe and thus the strain. This constitutes an additional edge in the graphical model, but has no significant impact on the inference process.
%

The updated digital state and quantities of interest are then used to compute the reward via the evaluation factor, $\phi_t^{\textrm{evaluation}}$ (Eq.~(\ref{eqn:phievaluation})). This reward factor is evaluated via the reward functions \eqref{eq:reward1}--\eqref{eq:reward3}.

The sequential Bayesian inference process proceeds as follows. When a new piece of observational data is acquired from the physical asset, we increment the current timestep $t_c$, and denote the new data by $o_{t_c}$. We add unobserved node $S_{t_c}$ and observed node $O_{t_c} = o_{t_c}$ to the graph. We also add the nodes $D_{t_c}$, $Q_{t_c}$, and $R_{t_c}$, if they do not already exist due to a past prediction. The edges connecting these nodes encode the probabilistic factors described in this section. With the graph assembled, we assimilate the newly acquired observational data in order to update our belief state, \eqref{eqn:factorization_tc}. This type of belief propagation is a classical task in probabilistic graphical models, and a wide variety of algorithms can be applied depending on graph structure and the nature of variables in the graph \cite{koller2009probabilistic}. In our demonstrative example, we are performing inference over a Bayesian network (directed, acyclic graph), and the sample spaces of all unobserved random variables are discrete (note that the quantities of interest and rewards are computed via continuous functions, but the set of possible inputs is the discrete set of states and controls). This allows us to use the classical sum-product algorithm to perform the belief update exactly. Once the joint belief state is computed we can perform marginalization to obtain posterior marginal distributions over any variable in the graph, for example, to estimate the current structural health of the UAV.
\paragraph*{Planning and Optimal Control}\label{sec:planning}
At each timestep the digital twin is tasked with responding to the evolving structural health intelligently by selecting and issuing a control input. This section describes how the proposed graphical model enables us to formulate a planning problem for which the solution is an optimal control policy. We approach the planning problem by first noting that the segment of the probabilistic graphical model (Figure~\ref{fig:graphicalmodel}) from the current timestep, $t_c$, until a chosen prediction timestep, $t_p$, can be viewed as a partially observable Markov decision process (POMDP) in which the state is a combination of the digital state and quantities of interest. In particular, we wish to choose control inputs, $u_{t_c},\ldots,u_{t_p}$, that steer the state of the asset in a way that maximizes the expected future reward.

The general solution to this planning problem is a control policy, $\pi$, of the form
\begin{linenomath}
\begin{equation}\label{eqn:policy}
u_{t} = \pi\left(p(D_{0},\ldots,D_{t}, Q_0,\ldots,Q_{t}  \mid  o_0,\ldots,o_{t}, u_0,\ldots,u_{t-1} )\right).
\end{equation}
\end{linenomath}
That is, the control policy maps from the current belief over the entire history to a control action. Thus, performing optimal control with a digital twin amounts to leveraging the models comprising the digital twin in order to find a control policy that maximizes the expected accumulated reward over the chosen prediction horizon. This can be stated as the optimization problem
\begin{linenomath}
\begin{equation}\label{eqn:planning}
\pi^* = \argmax_{\pi} \sum_{t = t_c+1}^{t_p} \gamma^{(t-t_c-1)} \mathbb{E}\left[R_t\right],
\end{equation}
\end{linenomath}
where $\gamma\in[0, 1]$ is a discount factor applied to reward at future timesteps.

The solution of these types of planning problems has been an active area of research in recent decades. Solving the problem exactly is typically intractable, but many effective approximate solution techniques exist. Recent approaches typically restrict focus to a subset of the belief space \cite{shani2013survey}, or adopt Monte Carlo sampling approaches \cite{silver2010monte} and learning algorithms \cite{ross2011bayesian,jaulmes2005active,karkus2017qmdp}. The choice of algorithm thus depends on the nature of the state, control, and observation spaces, as well as the desired frequency of replanning and the planning horizon. A key challenge in our demonstration is the continuous observation space (the space of possible strain measurements). We choose to make the planning problem tractable by approximating it as a \emph{fully-observable} Markov decision process (MDP). That is, we assume that in-flight our strain measurements will provide an accurate and certain estimate of the UAV structural health. This can be viewed as a way of decoupling the sensing and control problems: we design a controller assuming that our sensing capability is sufficient. We also use an infinite planning horizon, assuming that in practice the UAV would be flying successive missions indefinitely.

With these simplifying assumptions, we seek a policy of the form
\begin{linenomath}
\begin{equation}\label{eqn:policy_approx}
u_{t} = \tilde{\pi}\left(d^\star, q^\star\right),
\end{equation}
\end{linenomath}
where $d^\star$ and $q^\star$ are our best point estimates (we use the maximum a posteriori estimate) of the current state and quantities of interest for the UAV. This policy is suboptimal in general, but performs well on this demonstrative problem since our best point estimate of the UAV state is often accurate. We solve the simplified MDP planning problem offline for a policy of the form \eqref{eqn:policy_approx} via the classical value iteration algorithm \cite{russell2002artificial}. We use the reward function
\begin{linenomath}
\begin{equation}\label{eqn:reward}
  r_t(u_t, q_t) = r^{health}_t(q_t) + 2.5 r^{control}_t(u_t),
\end{equation}
\end{linenomath}
and a discount factor $\gamma=0.6$. This weighted reward is tuned to balance UAV aggression with self-preservation. Online, the policy is applied to the updated belief state \eqref{eqn:factorization_tc} at each timestep, in order to determine the next control input to apply.

%
%
%
%
%
%
%
%
%
%
%
%
%
%
%
%
%
%
\paragraph*{Extension to Prediction}
We extend the asset monitoring formulation to incorporate prediction over future timesteps. The prediction regime is represented by the segment of the probabilistic graphical model (Figure~\ref{fig:graphicalmodel}) from the current timestep, $t_c$, until a chosen prediction timestep, $t_p$. We can extend the target belief state to include prediction of digital state, quantity of interest, and reward variables up until the prediction horizon, $t_p$, as follows:
\begin{linenomath}
\begin{align}\label{eqn:factorization_tp}
& p(D_0,\ldots,D_{t_p},Q_0,\ldots,Q_{t_p},R_0,\ldots,R_{t_p}, \nonumber\\
& \hspace{10ex} U_{t_c+1},\ldots,U_{t_p} \mid o_0,\ldots,o_{t_c}, u_0,\ldots,u_{t_c})\nonumber\\
\propto&\prod_{t=0}^{t_p} \left[\phi_t^{\textrm{dynamics}}\phi_t^{\textrm{QoI}}\phi_t^{\textrm{evaluation}}\right]\prod_{t=0}^{t_c}\phi_t^{\textrm{assimilation}}\prod_{t=t_c+1}^{t_p} \phi_t^{\textrm{control}},
\end{align}
\end{linenomath}
where
\begin{linenomath}
\begin{equation}
\phi_t^{\textrm{control}} = p(U_t  \mid  D_t, Q_t).
\end{equation}
\end{linenomath}

The only additional term required is the control factor, $\phi_t^{\textrm{control}}$, which we define according to the control policy as
\begin{linenomath}
\begin{equation}
  p(u_t  \mid  d_t, q_t) = \left\{\begin{array}{ll}1 & \text{if } \tilde{\pi}(d_t,q_t) =  u_t,\\
  0 & \text{otherwise.}\end{array}\right.
\end{equation}
\end{linenomath}
Note that in the prediction regime, $t=t_c+1,\ldots,t_p$, we do not include a data assimilation factor and conditioning on $O_t=o_t$ is removed from the evaluation factor (i.e., we do not compute the $R^{error}_t$ term) as we have not yet observed data, $o_t$, for the future timesteps, $t>t_c$. With these minor adjustments and the additional control factor, both prediction and asset monitoring are performed seamlessly in a single pass of the sum-product algorithm.

%% file: sections/9_acknowledgments.tex
\section*{Data availability}
Experimental data acquired during the calibration experiments of this study are available in the public repository \texttt{UAV-experimental-calibration}\cite{michael_kapteyn_2021_4658935}.

\section*{Code availability}
The code used to perform the calibration procedure is available in the public repository \texttt{UAV-experimental-calibration}\cite{michael_kapteyn_2021_4658935}. This code, when combined with the provided experimental data, can be used to generate Figure \ref{fig:stiffness_results} and Figure \ref{fig:calibration_results_table}. Additionally, code used to implement the in-flight health monitoring simulation is provided in the public repository \texttt{UAV-digital-twin}\cite{michael_kapteyn_2021_4658878}. This simulation code was used to generate the results in Figure \ref{fig:online_results}.

The structural analysis software used to generate the results in this paper is Akselos Integra v4.5.9\footnote{https://akselos.com/}. Since the Akselos Integra software is proprietary and was used under license, we are unable to provide its source code. Instead, the model output data is provided directly in the repositories.

\section*{Acknowledgments}
This work was supported in part by AFOSR grant FA9550-16-1-0108 under the Dynamic Data Driven Application Systems Program, the SUTD-MIT International Design Center, the AFOSR MURI on managing multiple information sources of multi-physics systems awards FA9550-15-1-0038 and FA9550-18-1-0023, US Department of Energy grant DE-SC0021239, and the AEOLUS center under US Department of Energy Applied Mathematics MMICC award  DE-SC0019303.

\section*{Competing Interests}
Jessara Group sensors were used in the UAV experimental work described in this paper. Co-author Jacob Pretorius is a co-founder of Jessara. Purchase of the sensors for use in the research was reviewed and approved in compliance with all applicable MIT policies and procedures.